\documentclass[a4paper]{jpconf}
\usepackage[cp1252]{inputenc}
\usepackage{iopams}
\usepackage{graphicx}
\begin{document}
\title{The peculiarities of near-cathode processes in air discharge at atmospheric pressure}

\author{E~V~Parkevich$^{1,2}$, M~A~Medvedev$^{1,2}$, A~V~Agafonov$^{1}$, S~I~Tkachenko$^{2,3}$, A~V~Oginov$^{1}$, A~I~Khirianova$^{2}$, A~R~Mingaleev$^{1}$, T~A~Shelkovenko$^{1}$ and S~A~Pikuz$^{1}$}

\address{$^1$ Lebedev Physical Institute of the Russian Academy of Sciences, Leninsky Avenue 53, Moscow 119991, Russia}
\address{$^2$ Moscow Institute of Physics and Technology, Institutskiy Pereu\-lok 9, Dolgoprudny, Moscow Region 141700, Russia}
\address{$^3$ Joint Institute for High Temperatures of the Russian Acad\-emy of Sciences, Izhorskaya 13 Bldg~2, Moscow 125412, Russia}

\ead{parkevich@phystech.edu}

\begin{abstract}
Formation of near-cathode plasma at the instant of breakdown of the air gap was studied by the methods of picosecond laser probing. It was demonstrated that 1--2~ns after a sharp rise of the current through the discharge gap dense plasma clots with $N_{\rm e}\sim10^{20}$~cm$^{-3}$ and $\rmd N_{\rm e}/\rmd x \sim 10^{24}$~cm$^{-4}$ are formed on the cathode surface. It was shown that these highly ionized regions lead to initiation and development of the spark channel originating from the cathode. We propose that the observed formations correspond to erosive plasma formed from the cathode material.
\end{abstract}

\section{Introduction}

The spark discharge has been studied for more than a century \cite{bib_1, bib_2}. During the course of the studies in this research field it became clear that initiation of the breakdown of the discharge gap and subsequent appearance of the spark channels with high electron density is due to formation of micrometer-sized regions of intense plasma \cite{bib_3, bib_4}. Typical evolution time of these regions is very short (1~ns). The presence of such local areas of high ionization can lead to an increase in the power and energy input into these regions. Consequently, these areas greatly affect the parameters of the entire discharge. In this case near-electrode processes play a key role and are often accompanied by significant erosion of the electrodes. With employment of spectral methods it was demonstrated in \cite{bib_4, bib_5, bib_6} that the vapors of the electrode material can be present in the discharge gap. In a number of papers \cite{bib_7, bib_8, bib_9} the electron density of the spark channel plasma as high as $~10^{17} \textrm{--} 10^{18}$~cm$^{-3}$ was reported; as opposed to the case of streamers \cite{bib_10}, the spark channels were shown to originate from the electrode surface exclusively. This means that at an early stage of the spark channel development the key role belongs to erosive plasma. Such plasma exhibits high conductivity and facilitates a significant energy input into the developing spark channel \cite{bib_4}. Nevertheless, the mechanism of the electrode erosion at the moment of breakdown of the discharge gap and the role of the erosion products in further formation of the discharge are still not well-understood and require additional investigation.

In this paper, we attempt to trace the formation of near-cathode plasma at the instant of breakdown (with sharp increase of the current through the discharge gap) with high temporal and spatial resolution and to clarify the role of erosion plasma in formation of spark channels in air gaps at atmospheric pressure. Diagnostic methods based on picosecond laser probing were employed to obtain quantitative information on the near-cathode plasma.

\section{Experimental setup}

\begin{figure}[t]
\begin{center}
\includegraphics[width=1.0\columnwidth]{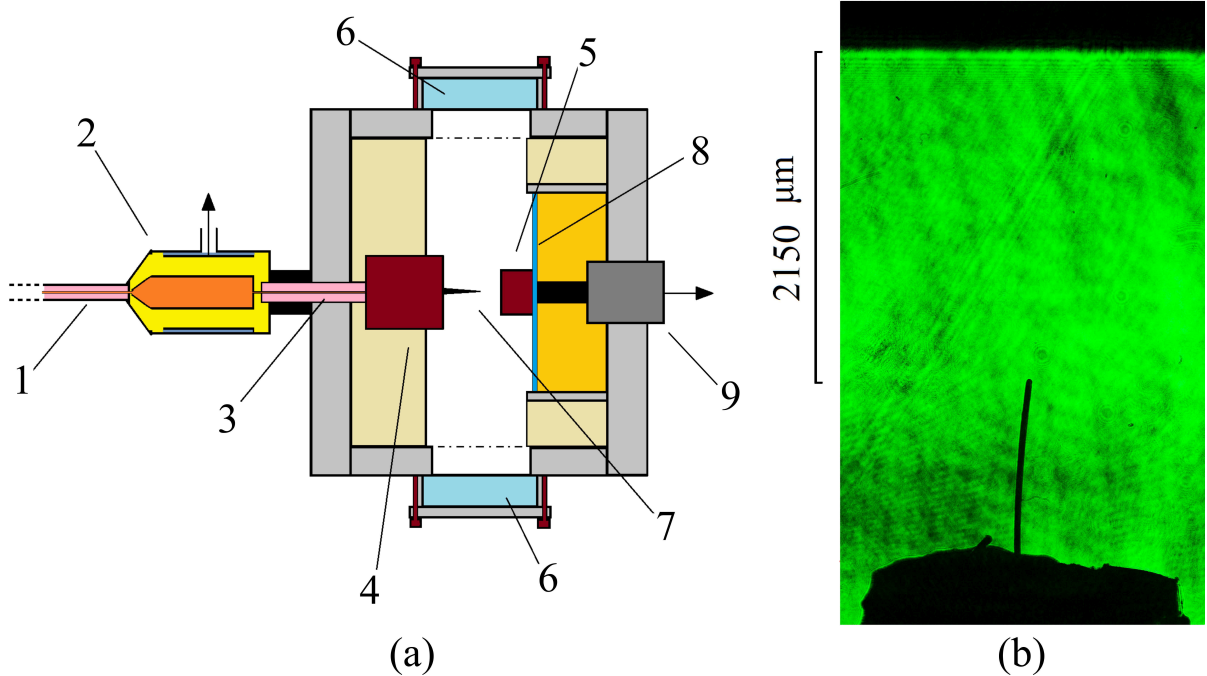}
\end{center}
\caption{\label{fig1}Scheme of discharge chamber with voltage and current detectors \pt(a): 1---transmitting line; 2---capacitive voltage divider; 3---transmission line with length of 2 cm; 4---cathode; 5---anode; 6---diagnostic windows (4); 7---discharge gap; 8---current shunt; 9---vacuum-tight coaxial BNC-connector. Shadowgram of discharge gap \pt(b). }
\end{figure}

\begin{figure}[t]
\begin{center}
\includegraphics[width=0.7\columnwidth]{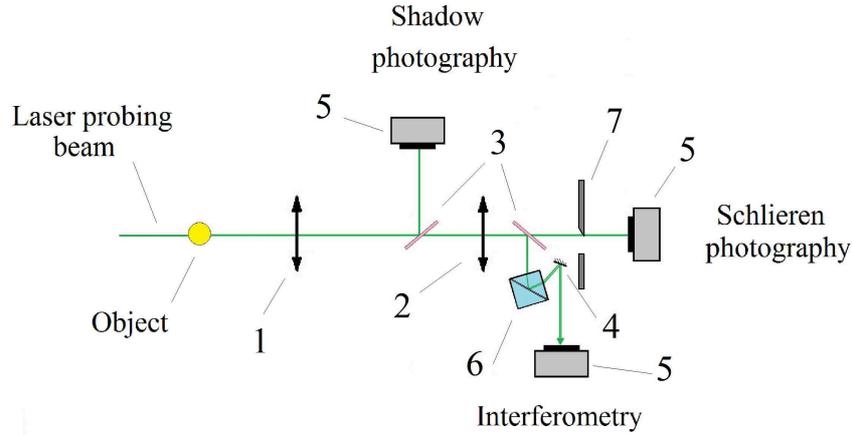}
\end{center}
\caption{\label{fig2}Setup for laser probing: 1{,\,}2---lenses with resolution of 500 and 400 line pairs/mm; 3---beam splitters; 4---rotary mirrors; 5---CCD; 6---air wedge shearing interferometer; 7---asymmetric schlieren diaphragm. }
\end{figure}

The investigations were carried out at the experimental setup with a high-voltage cable generator launched by an ignition laser beam with the jitter of $\sim 1$~ns. The generator provided output voltage pulses with their amplitude up to 20~kV (up to 40~kV in an open circuit) and pulse duration up to 30~ns with the rise time of $\approx 4$~ns and the maximum generator current of $\approx 500$~A. A Lotis LS-2151 Nd:YAG laser emitting at 1064~nm and 532~nm with the pulse energy up to 80~mJ was used. At 532~nm the pulse duration (full width at half magnitude, FWHM) was 70~ps. A part of the energy (approximately 90\%) of the output laser beam (both harmonics) was used to launch the high-voltage generator. The rest of the radiation (less than 10\%, the second harmonic only) was used for probing the discharge.

To implement the laser probing an optical system for simultaneous recording of interference, shadow and schlieren images of the discharge gap was developed. Figures~\ref{fig1} and \ref{fig2} show a schematic representation of the optical recording system, scheme of the discharge chamber as well as the shadowgram of the discharge gap. The spatial resolution of the system with 4x image magnification was $\approx 3$~$\mu$m ($\approx 70\%$ of the diffraction limit). The signals from the current and voltage sensors were recorded by a Tektronix TDS 3054B digital oscilloscopes with the bandwidth of 500~MHz. The moment of launching the high-voltage generator by the igniting laser beam was determined by a THORLABS-DET10A/M photo-detector and corresponded to the onset of the photo-detector signal increase. The temporal resolution of the photo-detector was 1~ns. The accuracy of the igniting laser beam registration was 0.5~ns with all optical delays, signal cable delays, the location of the current and voltage sensors as well as the sampling frequency of the oscilloscopes taken into account.

The air discharge at atmospheric pressure was studied using the ``pin-to-plane'' electrode geometry. The wires with the diameter of several tens of microns produced by ``California Fine Wires'' were used as the cathode. To achieve a better electrical contact the wire was soldered to a brass cone inserted into the cathode rod. The gap between the wire and the copper anode (with the diameter of 10~mm) was 2~mm, the distance between the cathode cone and the anode was 3~mm. The use of the point cathode provided both achievement of a strong field near the cathode and initiation of the discharge at a particular region. Moreover, the adjustment accuracy of the optical system focused at the object under study was improved.

\section{Experimental results and discussion}

\begin{figure}[t]
\begin{center}
\includegraphics[width=1.0\columnwidth]{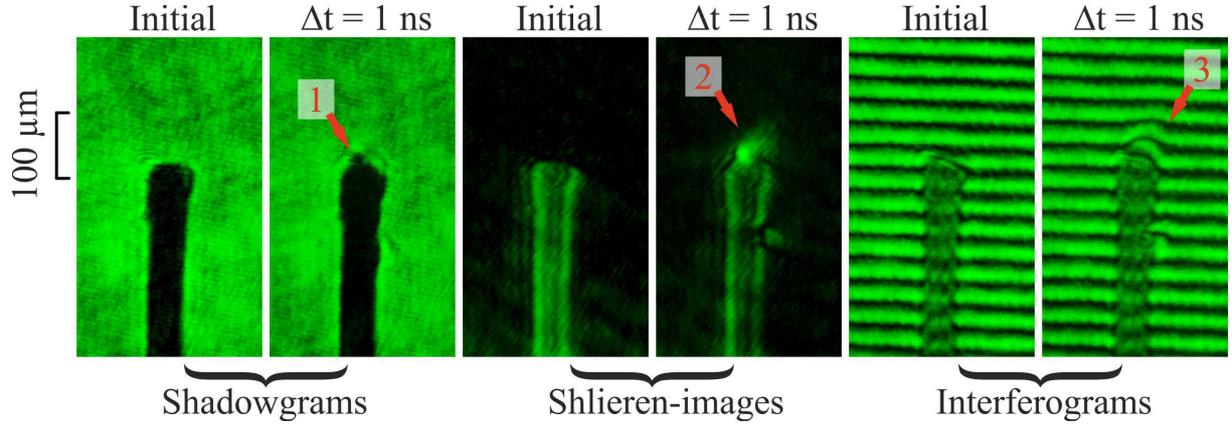}
\end{center}
\caption{\label{fig3}Development of anode-directed spark channel from the top of copper wire of diameter 50~$\mu$m. Photos were obtained after $\Delta t = 1$~ns from breakdown. Figures in photographs denote: 1---dense plasma clot; 2---areas of high ($\rmd N_{\rm e}/ \rmd x \sim 10^{22} \textrm{--} 10^{24}$~cm$^{-4}$) gradients of electron density; 3---developing spark channel. }
\end{figure}

Figure~\ref{fig3} shows typical photographs of the near-cathode region obtained before the discharge and those obtained 1~ns ($\Delta t$) after the gap breakdown ($t_{\rm br} \approx 3.5$~ns) demonstrating the development of the anode-directed spark channel originating from the cathode.

\begin{figure}[t]
\begin{center}
\includegraphics[width=1.0\columnwidth]{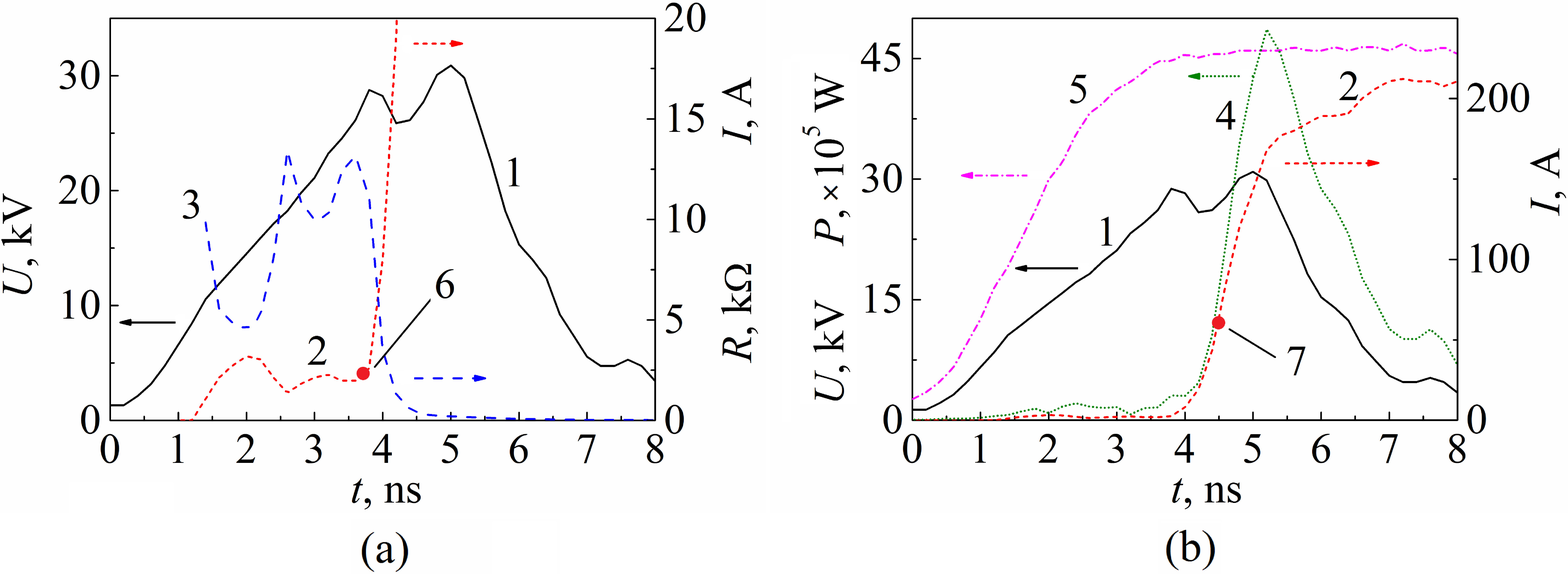}
\end{center}
\caption{\label{fig4}Discharge characteristics \pt(a, b): 1---gap voltage ($U$); 2---discharge current ($I$); 3---gap resistance ($R$); 4---electrical power input ($P$); 5---signal from voltage ($U$) sensor in open circuit; 6---the moment of breakdown ($t_{\rm br}$); 7---the moment of probing. Waveforms on the left figure \pt(a) show the behavior of the discharge current before the breakdown and its correlation with gap voltage and gap resistance. Waveforms on the right figure \pt(b) show the behavior of the discharge current after the breakdown and its correlation with drop in gap voltage as well as electrical power input.}
\end{figure}

\begin{figure}[t]
\begin{center}
\includegraphics[width=0.9\columnwidth]{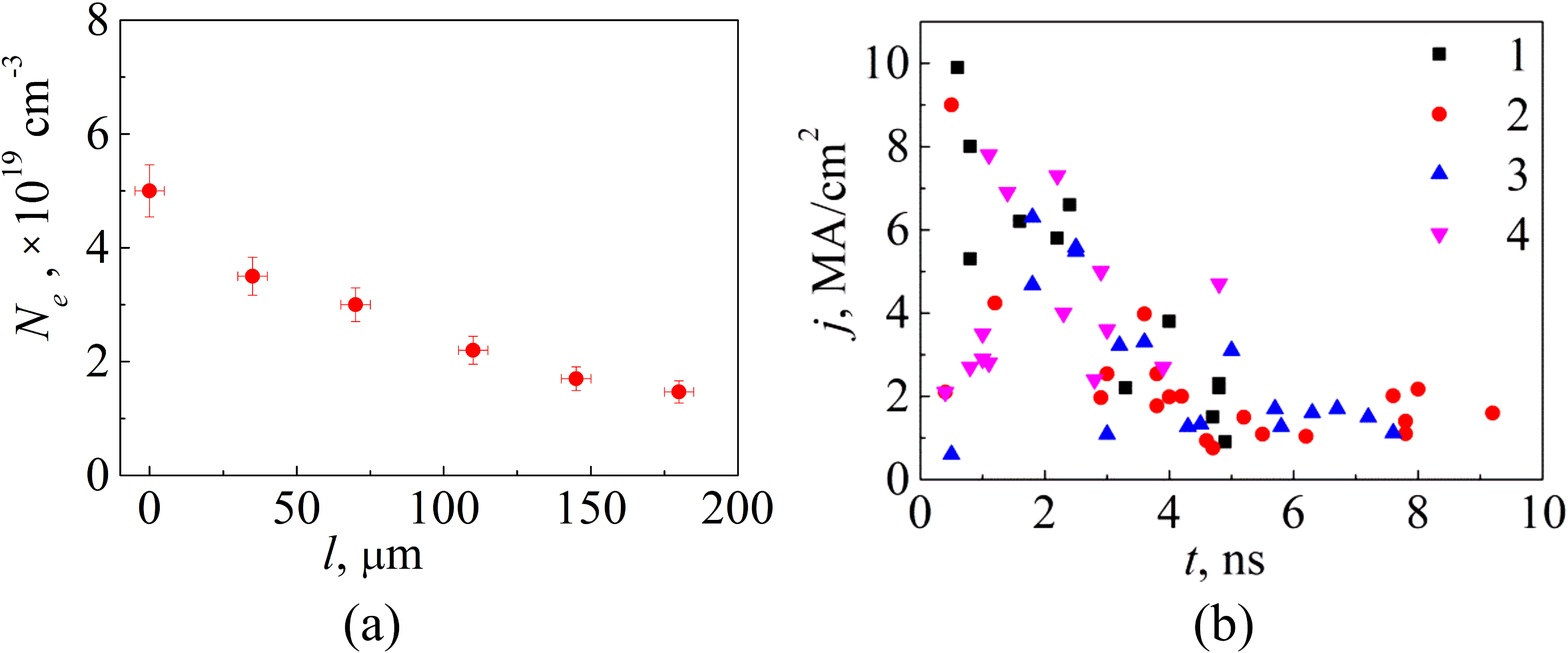}
\end{center}
\caption{\label{fig5}Typical distribution of electron density ($N_{\rm e}$) on the axis of spark channel with diameter of $\sim 100$~$\mu$m \pt(a) obtained after breakdown ($\Delta t = 4$~ns). Temporal dependence of current density ($j$) near wire end \pt(b): 1---gold wire of diameter 50 $\mu$m; 2---copper wire of 50 $\mu$m; 3---copper wire of 100 $\mu$m; 4---molybdenum wire of 40 $\mu$m. }
\end{figure}

It was established that approximately 1~ns after the breakdown clots of plasma appear at the top of the pin-cathode. These plasma regions have the size of 10~$\mu$m and are not transparent to optical radiation. Then, one or more highly ionized spark channels directed towards the anode are formed from these clots \cite{bib_11}. Within 1--1.5~ns after the breakdown the typical rise rate of the discharge current is $\rmd I/\rmd t \sim 100$~A/ns (figure~\ref{fig4}). The average rate of the channel expansion in the longitudinal direction is $70 \pm 5$~$\mu$m/ns, in the transverse direction the corresponding value is $30 \pm 5$~$\mu$m/ns. It was shown by interferometry, figure~\ref{fig5}\pt(a), that typical electron density in the developing spark channel with diameter of $\sim100$~$\mu$m can be as high as $N_{\rm e} = 5 \times 10^{19}$~cm$^{-3}$ \cite{bib_12, bib_13}. The accuracy of determining the diameter of the spark channel and the magnitude of the shift of the interference fringes (in units of the band shift) was about $\Delta d \approx 5$~$\mu$m and $\Delta k \approx 0.05$, respectively. The error in the determination of the electron density did not exceed 10\%. The processing method is described in detail in \cite{bib_14}. To calculate the electron density the expression for the refractive index of ideal plasma was used \cite{bib_15}:
\begin{equation}
n\approx 1-4.49\times 10^{-14} \lambda^{2}N_{e}.
\end{equation}

Here $n$ is the refractive index of the object, $\lambda = 532\times 10^{-7}$~cm is the wavelength of the probing radiation, $N_{\rm e}$~[cm$^{-3}$] is the electron density. The expression was derived on the assumption that the frequency of the probe radiation is well separated from all the resonance transitions of the particles present in the discharge and the frequency of electron-ion collisions is much lower than that of the probing radiation. Additional spectral analysis of the gas-discharge plasma formed in air at atmospheric pressure performed in the framework of the PrizmSPECT program \cite{bib_16} demonstrated that within $T_{\rm e} = 3\textrm{--}5$~eV temperature range the average degree of ionization is $z = 1$.

Figure~\ref{fig5}\pt(b) shows the typical temporal dependence of the current density in the spark channel near the surface of the wire end. The region with the size smaller than 50~$\mu$m was considered. The current density was estimated on the assumption that the discharge current detected by the shunt flows through the developing spark channel visible in the interferograms (see figure~\ref{fig3}). We regard only single homogeneous spark channels which are axisymmetric over their length. The data were obtained using the cathode wires made from different materials and having different diameters (Au, 50 $\mu$m; Cu, 50 and 100 $\mu$m; Mo, 40 $\mu$m). Therefore, the samples have different thermal and electrical properties. By the time a visible clot of plasma appears the current density in the clot is $j \sim 10^{5} \textrm{--} 10^{6}$~A/cm$^{2}$, and 1--1.5~ns after the breakdown this value can be as high as $j \sim 10^{7}$~A/cm$^{2}$. After 2--3~ns from the breakdown the current density in the developing spark channel is $j \sim 10^{6}$~A/cm$^{2}$ and it is approximately constant in time.

The maximum of the current density obtained approximately 1~ns after the breakdown is associated with the features of the rise of the discharge current within the considered time interval. During this period, the channel expands in the transverse direction by only a few times, and the current increases by 2 orders of magnitude. At the time of the breakdown the discharge current is on average 2~A and the gap resistance is $R \sim 10$~k$\Omega$. The breakdown voltage is on average $U_{\rm br} = 20 \textrm{--} 30$~kV and is lower than the maximum achievable value of 40~kV. It is noteworthy that 1~ns after the breakdown the gap resistance drops from $R \sim 10 \textrm{--} 15$~k$\Omega$ to hundreds of Ohms with $R = (U-L \rmd I/ \rmd t)/I$, where $L \approx 10$~nH is the inductance of the cathode wire. At the same time, the voltage drop in the gap does not occur immediately and the corresponding decay time is on average 3--4~ns. Within the same time interval, the electrical power input $P = UI$ into the gap reaches several megawatts, see figure~\ref{fig4}\pt(b). This fact indicates that highly intensive ionization processes possibly occur in the discharge gap.

\begin{figure}[t]
\begin{center}
\includegraphics[width=1.0\columnwidth]{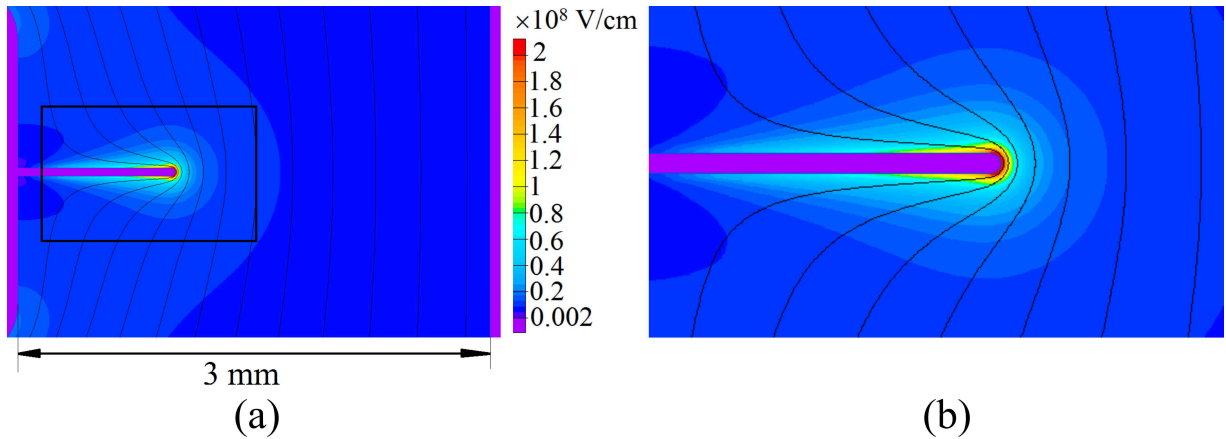}
\end{center}
\caption{\label{fig6}Distribution of electric field \pt(a) in discharge gap at breakdown voltage ($U_{\rm br}$) of 25~kV. Region \pt(b) of high electric field $E > 10^{6}$~V/cm. Step of isolines is 2~kV. The cathode wire is 50~$\mu$m in diameter. }
\end{figure}

Numerical simulation of the electric field distribution (figure~\ref{fig6}) near the surface of the wire end at the instant of the breakdown showed that by this moment the field can be as high as $E \sim 10^{7} \textrm{--} 10^{8}$~V/cm. The field was simulated in the electrostatic approximation by the finite element method using ELCUT-5.8 software \cite{bib_17} with the geometric parameters of the coaxial discharge chamber taken into account. We note that the real value of the macrofield near the wire end at the instant of the breakdown can be different from the obtained estimation. For example, this can be due to the effect of the space charges near the cathode surface. Nevertheless, such strong fields are achieved near different microinhomogeneities of the wire surface, where the field gain is 10--10$^{3}$. With such parameters of the electric field and current density the process of electron emission from metal can be of explosive character \cite{bib_18}. In this case the process will be accompanied by noticeable erosion of the cathode surface.

In the shadowgrams (see figure~\ref{fig3}) erosion plasma appears as opaque regions with the size of about 10~$\mu$m. This effect can be associated with both strong absorption of the probing radiation and refraction. It should also be noted that at the wavelength of 532 nm the critical value of the electron density $N_{\rm e}^{\rm c}$ is:
\begin{equation}
N_{\rm e}^{\rm c} = \frac{\pi m_{\rm e} c}{e^{2} \lambda ^{2}}\approx 3.9\times 10^{21}~[\rm cm^{-3}].
\end{equation}

Here $m_{\rm e}$ and $e$ are the electron mass and charge, $c$~[cm/s] is the speed of light in vacuum, $\lambda$~[cm] is the wavelength of the probing radiation.

For quantitative analysis of the formation processes of the observed dense near-cathode plasma, an important parameter is its electron temperature. However, on such a small scale it is difficult to obtain the corresponding information with subnanosecond temporal resolution by a direct method. Assuming that there are no significant differences in the processes of cathode explosion in air and in vacuum, it is possible to use the experimental data on the formation mechanisms of dense erosion plasma in a vacuum diode. As shown by numerous studies of erosion plasma in vacuum \cite{bib_19}, the temperature and the ionization degree of the plasma have the values of $T_{\rm e} = 1\textrm{--}5$~eV and $z = N_{\rm e}/N_{\rm i} =1 \textrm{--} 3$. These results provide a means for a quantitative analysis of the intensity attenuation of the probe radiation. The analysis can be performed in accordance with the Bouguer's law: $I_{\rm out}/I_{\rm in} = \exp(-\Theta_{\rm \nu}d)$, where $I_{\rm in}$ and $I_{\rm out}$ are the intensities of the incident and transmitted beam, $\rm d$ is the optical path length, and $\rm \Theta _{\rm \nu}$ is the absorption coefficient. In the approximation of the semiclassical theory of bremsstrahlung absorption \cite{bib_20}, the spectral absorption coefficient is given by the following expression (which takes the reradiation in the plasma volume into account \cite{bib_21}):
\begin{equation}
\Theta_{\rm \nu} = C_{1} \frac{z^{2}gN_{\rm e}N_{\rm i}}{T_{\rm e}^{1/2} \nu ^{3}} \Biggl [1-\exp \biggl(-\frac{h \nu}{k T_{\rm e}} \biggr) \Biggr].
\end{equation}

\begin{figure}[t]
\begin{center}
\includegraphics[width=0.5\columnwidth]{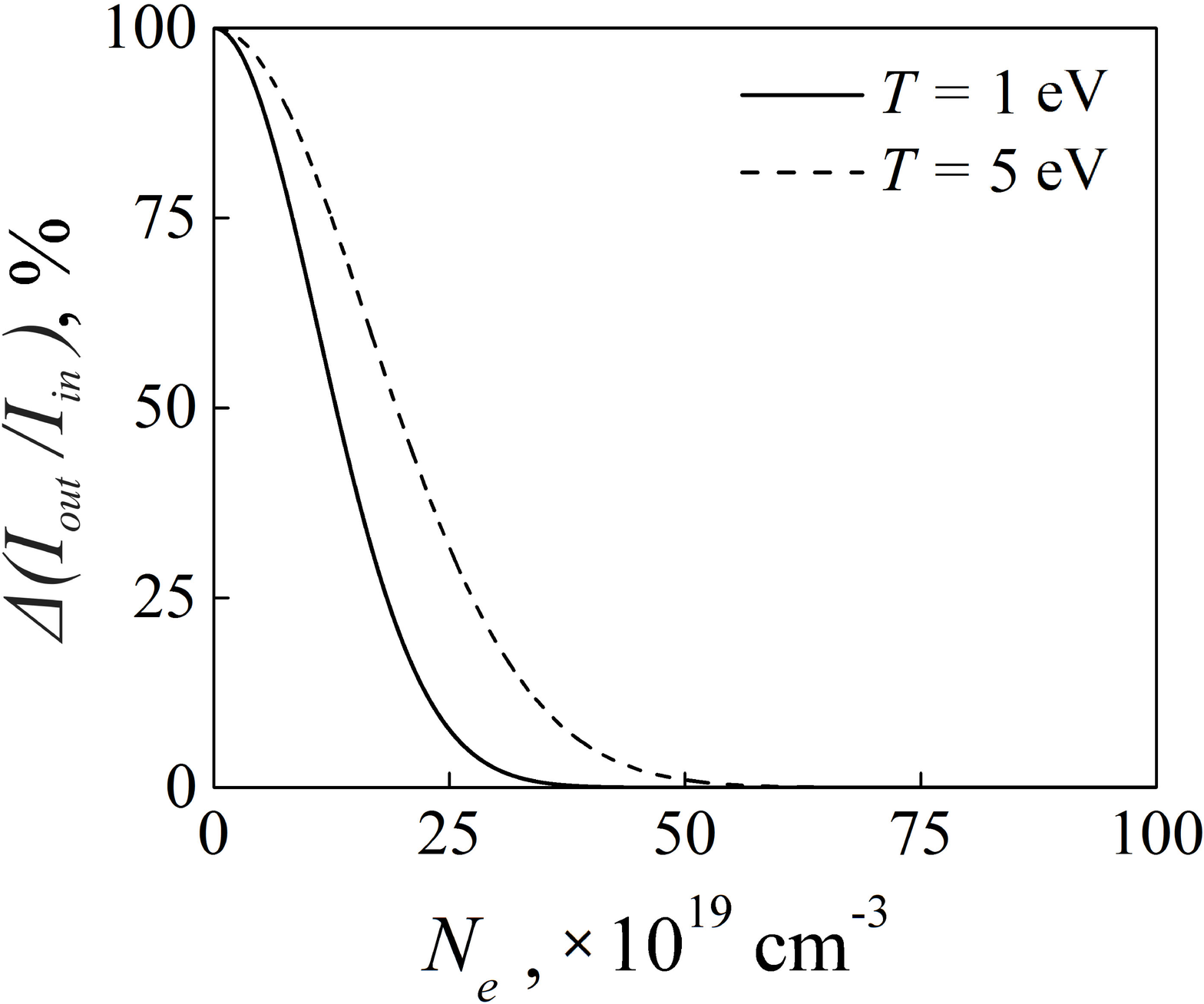}
\end{center}
\caption{\label{fig7}Dependence of relative change in probing beam intensity $\Delta (I_{\rm out}/I_{\rm in})$ on electron density ($N_{\rm e}$) at electron temperature $T_{\rm e} = 1$ and $5$~eV, average ion charge $z = 1$ and size of observed object $d = 10$~$\mu$m. }
\end{figure}

Here $C_{1} = 3.69\times 10^{8}$~cm$^{5}$K$^{1/2}$c$^{-3}$, $z$ is the average ionization degree, $N_{\rm e}$ and $N_{\rm i}$ are the electron and ion densities ($N_{\rm e} = zN_{\rm i}$~for~$z > 1$), $T_{\rm e}$ is the electron temperature, $h \nu$ is the photon energy of the probing radiation, and $g$ is the Gaunt factor (assumed to be 2). Figure~\ref{fig7} shows the dependence of the relative change in the probing radiation intensity $ \Delta (I_{\rm out}/I_{\rm in})$ on $N_{\rm e}$, $T_{\rm e}$, $z$. The electron density was assumed to be $N_{\rm e} = (1 \textrm{--} 100) \times 10^{19}$~cm$^{-3}$, the average ion charge was $z = 1$, the electron temperature was $T_{\rm e} = 1$ and $5$~eV, and the optical path length was $d = 10$~$\mu$m. Estimates show that more than 90\% absorption of the probing beam was achieved at $\langle N_{\rm e} \rangle =3\times 10^{20}$~cm$^{-3}$.

It should be noted that the cathodes made from different materials, namely, copper, gold, molybdenum, and stainless steel, provided observation of the clots of opaque plasma (see figure~\ref{fig3}). In this case, it is unlikely that the resonance absorption occurs in a clot of dense plasma. In addition, the line width of the laser radiation at 532~nm is 0.07~nm. Therefore, one can assume that the laser radiation frequency is well separated from the frequencies of all the resonant transitions of the particles present in the discharge, both near the cathode and in the entire discharge gap.

According to the calculations are given in \cite{bib_22} the deflection angle of the probing beam can be as high as 14$^{\circ}$. The estimates were made for clot with size of $d\sim 10$~$\mu$m with the electron density of $N_{\rm e} \sim 10^{20}\textrm{--}10^{21}$~cm$^{-3}$ and typical density gradient of $\rmd N_{\rm e}/\rmd x \sim 10^{24}$~cm$^{-4}$. It should be noted that the obtained value of the refraction angle exceeds the aperture angle ($\approx 11^{\circ}$) of the first lens in the optical recording system.

Thus, the dense plasma observed in the recorded images (see figure~\ref{fig3}) can result from both absorption and strong refraction of the probing radiation provided that the typical size of the clots is $d\sim10$~$\mu$m and the corresponding electron density is $N_{\rm e}\sim10^{20} \textrm{--} 10^{21}$~cm$^{-3}$. It should be noted that in the paper only qualitative estimations were made and a more detailed analysis of the refractive index of the objects under study is required.

Analytical and numerical analysis of the spark channel evolution after the breakdown is a complex and nonlinear problem. However, it is possible to give a qualitative explanation to the mechanism of the plasma formation on the basis of the data presented in the paper.

The appearance of the dense plasma clots can be accounted for in terms of explosive electron emission (EEE). In this case, the fields needed to produce such plasma in air are lower than those required in vacuum as the plasma obtained by gas ionization near the cathode emitter stimulates the EEE \cite{bib_4}. If the applied electric field is sufficient for cold emission to result in the appearance of at least one electron, then the field at the cathode is increased due to impact ionization and diffusion of positive ions towards the cathode. Consequently, the electron current from the cathode will be increased alongside with the ion current towards the cathode, providing an increase in the electric field at the cathode. These processes last until the microinhomogeneities on the wire end surface explode. It should also be noted that the EEE can be additionally enhanced by various dielectric films present on the cathode surface.

Therefore, micron-scale clots of dense erosive plasma are formed within the near-cathode region. Further development of the considered processes will result in initiation of a spark channel as the highly ionized regions formed around the wire end will propagate towards the anode with the rates higher than the propagation rates for the dense erosion plasma.

In conclusion, it should be noted that behaviour of the discharge observed with different geometries of the electrodes was similar to the results described above.

\section{Conclusion}

In this paper the evolution of the spark channel during the first nanoseconds after the breakdown was studied with high temporal and spatial resolution using laser probing methods. Opaque dense plasma clots with the size of about 10~$\mu$m were observed near the cathode surface approximately 1~ns after the breakdown. These clots are most likely the erosion plasma formed from the cathode material. Analysis of the absorption and refraction of the laser radiation demonstrated that the observed formations can have the electron density greater than 10$^{20}$~cm$^{-3}$ and typical density gradients of the order of $10^{22} \textrm{--} 10^{24}$~cm$^{-4}$. The obtained data indicate that the formation of the spark channel at the cathode surface can be associated with explosive electron emission.

It should also be noted that the obtained data on the discharge gap breakdown are important for both development of the models regarding the conductivity of the spark channel and refinement of the existing ones.

\ack
The experimental study was supported by the Russian Foundation for Basic Research (grant No.\,18-32-00566). The development of the optical recording system was partially supported by the Russian Foundation for Basic Research (grants No.\,17-08-01690;18-02-00631). The plasma analysis was supported by the agreement with Cornell University, Electrical and Computer Engineering School, under Prime Agreement DE-NA0003764 from National Nuclear Security Administration DOE.

\section*{References}
\bibliographystyle{iopart-num}
\bibliography{article_367_v6}

\providecommand{\newblock}{}
\begin{thebibliography}{10}
\expandafter\ifx\csname url\endcsname\relax
  \def\url#1{{\tt #1}}\fi
\expandafter\ifx\csname urlprefix\endcsname\relax\def\urlprefix{URL }\fi
\providecommand{\eprint}[2][]{\url{#2}}

\bibitem{bib_1}
Walters J~P 1969 {\em Appl. Spectrosc.\/} {\bf 23}(4) 317--31

\bibitem{bib_2}
Bazelyan E~M and Raizer {\relax Yu}~P 2000 {\em Lightning Physics and Lightning
  Protection\/} (UK: IOP Publishing Ltd.)

\bibitem{bib_3}
Raizer {\relax Yu}~P 1991 {\em Gas Discharge Physics\/} (Berlin:
  Springer-Verlag)

\bibitem{bib_4}
Korolev {\relax Yu}~D and Mesyats G~A 1998 {\em Physics of Pulsed Breakdown in
  Gases\/} (Ekaterinburg: Ural Division of the Russian Academy of Science)

\bibitem{bib_5}
Andreev S~I and Novikova G~M 1976 {\em Opt. Spectrosc.\/} {\bf 40}(2) 130--4

\bibitem{bib_6}
Andreev S~I and Novikova G~M 1975 {\em Zh. Tekh. Fiz.\/} {\bf 45}(8) 1692--1703

\bibitem{bib_7}
Kozyrev A~V, Korolev {\relax Yu}~D and Tinchurin K~A 1988 {\em Fiz. Plazmy\/}
  {\bf 18}(8) 1003--1007

\bibitem{bib_8}
{van der Horst} R~M, Verreycken T, {van Veldhuizen} E~M and Bruggeman P~J 2012
  {\em J. Phys. D: Appl. Phys.\/} {\bf 45} 345241

\bibitem{bib_9}
Lo A, Cessou A, Lacour C, Lecordier B, Boubert P, Xu D~A, Laux C~O and Vervisch
  P 2017 {\em Plasma Sources Sci. Technol.\/} {\bf 26} 045012

\bibitem{bib_10}
Raether H 1964 {\em Electron Avalanches and Breakdown in Gases\/} (London:
  Butterworths)

\bibitem{bib_11}
Oginov A~V, Parkevich E~V, Shpakov K~V, Rodionov A~A, Baidin I~S, Agafonov A~V
  and Medvedev M~A 2018 {\em J. Phys.: Conf. Ser.\/}  This issue

\bibitem{bib_12}
Parkevich E~V, Tkachenko S~I, Agafonov A~V, Mingaleev A~R, Romanova V~M,
  Shelkovenko T~A and Pikuz S~A 2017 {\em J. Exp. Theor. Phys.\/} {\bf 124}
  531--9

\bibitem{bib_13}
Parkevich E~V, Khirianova A~N, Agafonov A~V, Tkachenko S~I, Mingaleev A~R,
  Shelkovenko T~A, Oginov A~V and Pikuz S~A 2018 {\em J. Exp. Theor. Phys.\/}
  {\bf 126} 423--30

\bibitem{bib_14}
Khirianova A~I, Parkevich E~V and Tkachenko S~I 2018 {\em J. Phys.: Conf.
  Ser.\/}  This issue

\bibitem{bib_15}
Alpher R~A and White D~R 1965 {\em Optical Interferometry\/} (New York:
  Academic Press) pp 431--73

\bibitem{bib_16}
Http://www.prism-cs.com/Software/PrismSPECT/overview.html

\bibitem{bib_17}
Http://elcut.ru/

\bibitem{bib_18}
Mesyats G~A 2011 {\em Explosive Electron Emission\/} (Moscow: FIZMATLIT)

\bibitem{bib_19}
Proskurovskii D~I 2010 {\em Emissionnaya Elektronika\/} (Tomsk: Tomsk State
  University)

\bibitem{bib_20}
Ochkin V~N 2009 {\em Spectroscopy of Low Temperature Plasma\/} (Weinheim:
  Wiley)

\bibitem{bib_21}
Zel'dovich {\relax Ya}~B and Raizer {\relax Yu}~P 1968 {\em Elements of Gas
  Dynamics and the Classical Theory of Shock Waves\/} (New York: Academic)

\bibitem{bib_22}
Dushin L~A (ed) 1973 {\em Zondirovanie Neodnorodnoi Plazmi Elektromagnitnimi
  Volnami\/} (Moscow: Atomizdat)

\end{thebibliography}

\end{document}